\documentclass[showpacs,amssymb, amsmath,nobibnotes, aps,prl,secnumarabic
,twocolumn%
]{revtex4}

\usepackage{bm}
\usepackage{natbib}
\usepackage{graphicx}

\begin{document}

\title {Strong ExB shear flows in the transport barrier region in H mode plasma}
\newcommand{\iotabar}{\raisebox{-1pt}{$\mathchar'40$}\mkern-5.43mu\iota}

\author{H. Xia}
\email{hua.xia@anu.edu.au}
\author{M. G. Shats}
\email{Michael.Shats@anu.edu.au}
\author{H. Punzmann}
\email{Horst.Punzmann@anu.edu.au}
\affiliation{Plasma Research Laboratory, Research School of
Physical Sciences and Engineering,
\\Australian National University, Canberra ACT 0200, Australia}
\date{\today}%

\begin{abstract}

We report the first experimental observation of stationary zonal
flow in the transport barrier region of the H mode plasma. Strong
peaks in $E_r$ shear mark the width of this region. Strong $m=n=0$
low-frequency ($f<$ 0.6 kHz) zonal flow is observed in regions of
increased $E_r$, suggesting substantial contribution of zonal flow
to the spatial modulation of $E_r$ radial profiles. Radial
localization of the zonal flow is correlated with a region of zero
magnetic shear and low-order (7/5) rational surfaces.

\end{abstract}


\maketitle

Transport barriers (TBs) are radially localized regions in
toroidal plasma where radial transport of particles or energy is
drastically reduced. In the high confinement mode (H mode)
\cite{Wagner1982}, the presence of a TB is manifested as a steep
density (or temperature) gradient near the plasma boundary. The
top of this region is sometimes referred to as a pedestal.

Characteristics of the H mode edge TBs are important. Spatial
structure of a TB is closely related to the global stability,
confinement and the plasma performance (for a review see, for
example, \cite{Fujisawa_PPCF_2003}). Understanding and predicting
characteristics of the TBs has become a focus of the international
fusion community (see, e.g., \cite{Hatae2001,Fujita2002}). The
ultimate goal of these studies is the optimization of the radial
profiles of the plasma parameters in the future fusion reactor
\cite{ITER}.

The formation of a TB has been ascribed to the generation of a
sheared radial electric field (or \mbox{\textbf{E} $\times $
\textbf{B}} flow, where \textbf{E} is the electric field and
\textbf{B} is the magnetic field) which leads to the reduction in
turbulence and transport \cite{Terry2000, Gohil_PPCF_2002,
Hahm_PPCF_2002}. However, the physics of the TB formation is not
yet well understood. Experimental studies of TBs are restricted
due to difficulties in measuring radial parameter profiles with
sufficient spatial and temporal resolution.

In this Letter we report detailed experimental studies of the TB
structure in H mode of the H-1 heliac. It is shown for the first
time, that distinct features in the electron density profile,
marking the pedestal and the foot of TB, spatially coincide with
radially localized strongly sheared \mbox{\textbf{E} $\times $
\textbf{B}} flows. These radial regions are also identified as
regions where strong stationary $m = n = 0$ zonal flows are
localized in H mode. The radial localization of zonal flows also
coincides with the position of a low-order rational surface and a
minimum in the magnetic shear. These results confirm, to some
extent, a hypothesis based on results of the gyrokinetic
simulations that strong zonal flows developing near rational
surfaces can provide a trigger for the TB formation
\cite{Waltz2006}.

We present results obtained in the H-1 toroidal heliac
\cite{Ham90} (major radius of $R$~=~1~m and mean minor plasma
radius of about $\left\langle a \right\rangle\approx 0.2\ $ m)
under the following plasma conditions (see, for example,
\cite{Sha02a} and references therein):
$n_e=1\times10^{18}$~m$^{-3}$, $T_{e} \sim 10$~eV, $T_{i} \sim
40$~eV in argon at filling pressure of $(1 - 4)\times
$10$^{-5}$~Torr and at low magnetic fields, $B = (0.05 - 0.12)$~T.
Such plasma is produced by $\sim $~80~kW of the radio-frequency
waves at 7~MHz. Several combinations of Langmuir probes (single,
triple probes) are used to characterize plasma parameters, such as
the electron density, electron temperature, and electrostatic
potential, as described in \cite{Sha02b}. Probes are also used to
characterize poloidal and toroidal wave numbers of turbulent
fluctuations \cite{Shats_PPCF_06}.

\begin{figure}
\includegraphics{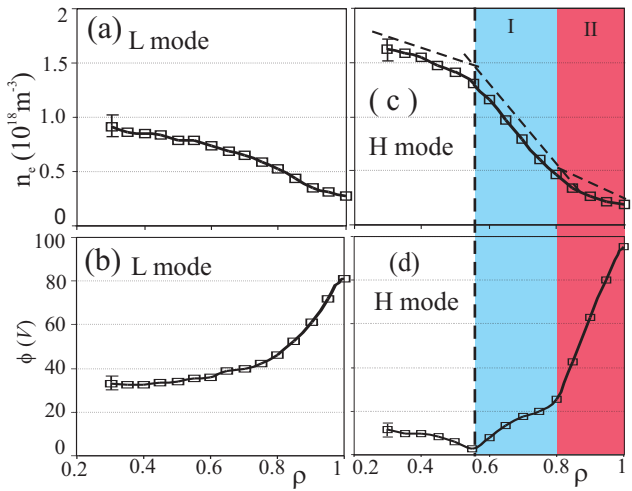}
\caption{\label{fig1} Radial profiles of (a) electron density, (b)
plasma potential in L mode, and (c) electron density, (d) plasma
potential in H mode, respectively. Dashed guide lines and shading
are used to mark two radial regions in the transport barrier (I
and II).}
\end{figure}

The high confinement mode observed in H-1 \cite{Sha99} is similar
to H mode in tokamaks. Typical electron density and plasma
potential profiles are illustrated in Fig.~\ref{fig1} for low (L)
and H modes. L and H modes are achieved respectively above and
below critical magnetic field \cite{Sha99}. When the magnetic
field is close to the critical value, spontaneous L-H transitions
are observed. In these discharges, a triple probe is used to
measure the electron density and potential on a shot-to-shot
basis. Excellent reproducibility of the measurements allows
reliable determination of profiles without perturbing plasma by
the probe arrays.

Despite large differences in electron temperature, density, and
magnetic field, plasmas in H-1 and in the TB regions of large
tokamaks, are dimensionally similar. This dimensional similarity
has been discussed in \cite{Punzmann2004}, where it has been shown
that the width of TB measured in ion gyroradii is very similar to
that in, for example, DIII-D tokamak. However in absolute units,
the TB width in H-1 is substantially broader (30-40 mm) than that
in larger experiments with stronger magnetic fields and lighter
ions. This, in combination with low electron temperature in H-1,
opens an opportunity to study structure of the TB using probes
with sufficiently high spatial resolution.

The development of the TB in H mode plasma in H-1 is illustrated
in Fig.~\ref{fig1}. Radial profiles of the electron density and
the plasma potential in L mode are rather featureless as seen in
Fig.~\ref{fig1}(a,b). In H mode, the central density doubles while
the plasma potential becomes more negative in the central region
and more positive at the edge (Fig.~\ref{fig1}(c,d)). The increase
in the density coincides with the formation of the characteristic
kink in the density profile at about $\rho = r/a \approx 0.6$,
referred to as the pedestal. The $n_e$ profile outside the
pedestal can be approximated by a straight line
(Fig.~\ref{fig1}(c)). The profile of the plasma potential $\phi$
also shows two characteristic kinks: one at the top of the TB, and
the other at $\rho = 0.8$, which we will refer to as the foot of
the TB. The third kink in the plasma potential is seen near the
last closed flux surface ($\rho = 1.0$) and is due to the reversal
of the radial electric field from negative (inside) to positive
(outside). We use dashed guide lines and shading throughout the
paper to mark two radial regions of interest: (I) - a region
between the top and foot of the TB, and (II) - a region between
the foot and the last closed flux surface.

\begin{figure}
\includegraphics{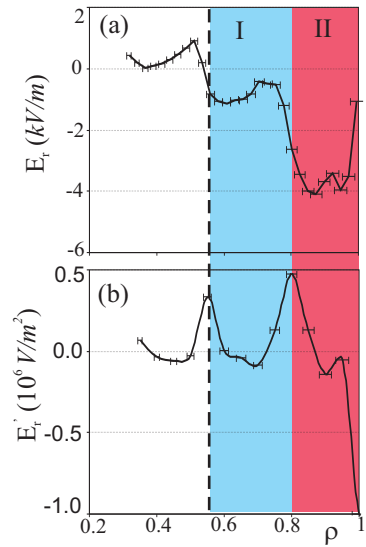}
\caption{\label{fig2} Radial profiles of (a) radial electric field
and (d) shear in the radial electric field, computed using plasma
potential profile of Fig.~\ref{fig1}(d).}
\end{figure}

Profiles of the radial electric field $E_r$ and its shear
$E_r^{\prime}$, derived from the radial profile of the plasma
potential are shown in Fig.~\ref{fig2}. Since $E_r$ is computed by
differentiating radial profile of the plasma potential, the
(negative) maxima of the radial electric field can not be
determined exactly. Three $E_r$ regions are seen: slightly
positive $E_r$ inside the top of the transport barrier,
substantial negative $E_r \approx$ -1 kV/m in region I, and even
more negative $E_r \approx$ -4 kV/m in region II. Correspondingly,
the $E_r^{\prime}$ has distinct peak at the top, at the foot of
the transport barrier and at the last closed flux surface.

\begin{figure}
\includegraphics{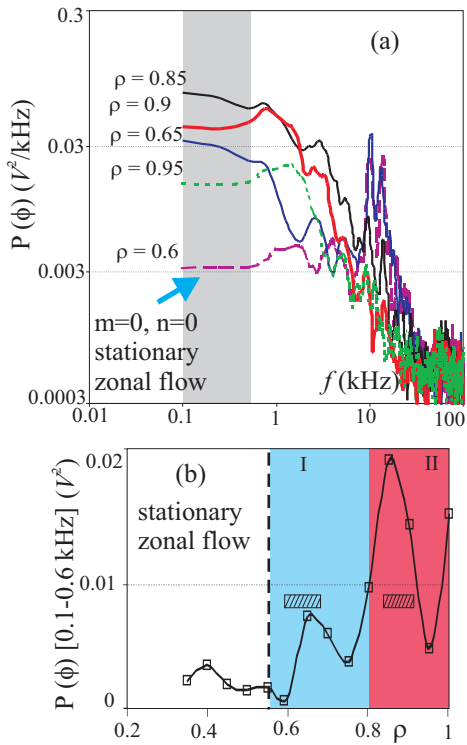}
\caption{\label{fig3}  (a) Power spectra of the plasma potential
in different radial regions; (b) radial profile of the spectral
power density of stationary zonal flows (0.1 $\sim$ 0.6 kHz).
Hatched boxes indicate radial positions of the $E_r$ maxima in
regions I and II. }
\end{figure}

Fluctuations in the electron density and potential are strongly
reduced in a broad range of frequencies from L to H mode as
discussed in \cite{Shats_PRE_05}. However, the low frequency
($f<$~0.6~kHz) spectral feature increases in some radial regions,
which will be discussed later. The power spectra of the
fluctuations in the plasma potential, $P(\phi)$, at various radial
positions in H mode are shown in Fig.~\ref{fig3}(a). The
low-frequency feature ($0.1 - 0.6$ kHz) is dominant in H mode.
Poloidal wave number $k_{\theta}$ of this low frequency component
is measured using two poloidally separated probes. Measured
poloidal wave number of $k_{\theta}= (2 - 5)~ $~m$^{-1}$ at
$f=(0.1 - 0.6)$~kHz is indicative of the mode number $m=0$. The
toroidal mode number is estimated using toroidally separated
probes, as described in \cite{Shats_PPCF_06}, and shows $n=0$.
Hence the strong low frequency fluctuations in the plasma
potential are identified as stationary zonal flows.  It should be
noted that it is usually difficult to align toroidally separated
probes to exactly the same poloidal position. As a result, a phase
shift between toroidally separated probes will occur due to the
uncertainty in the poloidal separation between the probes, $\Delta
y$:
\begin{equation}\label{mode_number_measurement}
\Delta \varphi(f) = k_{\|}(f)\Delta L_{\|} + k_{\theta}(f)\Delta y,
\end{equation}
\noindent where $\Delta L_{\|}$ and $\Delta y$ are toroidal and
poloidal separation between the probes respectively, and
$k_{\theta}(f)$ is known from the phase difference between the
probes which are poloidally separated. In case of a zonal flow, $m
= 0$, the second term on the right-hand side becomes zero (since
$k_{\theta} = 0$), such that the poloidal uncertainty $\Delta y$
becomes unimportant and the toroidal wave number can be reliably
estimated by measuring $\Delta \varphi$.

Spectra similar to those in Fig.~\ref{fig3}(a) have also been
observed in the Compact Helical System (CHS) using heavy-ion-beam
probe \cite{Fujisawa_PRL_04}. In that experiment, low frequency
potential structures were also identified as stationary zonal
flows.

The spectral power density of the zonal flow (shaded spectral
region of $f=(0.1 - 0.6)$ kHz in Fig.~\ref{fig3}(a)) varies along
the radius. Radial profile of the spectral power density of the
zonal flow in H mode is presented in Fig.~\ref{fig3}(b).
Stationary zonal flow is a band-like structure localized in radial
region of $0.6 < \rho < 1.0$.

Two hatched boxes drawn in Fig.~\ref{fig3}(b) indicate the
uncertainty in the radial positions of the (negative) $E_r$ maxima
in regions I and II ( Fig.~\ref{fig2}(a)). It can be seen that the
zonal flow maximum spatially coincides with the maximum in
(negative) $E_r$. This suggests that stationary zonal flow
directly contributes to mean $E_r$ and may be responsible for the
"corrugation" of the $E_r$ profile seen in Fig.~\ref{fig2}(a).

\begin{figure}
\includegraphics{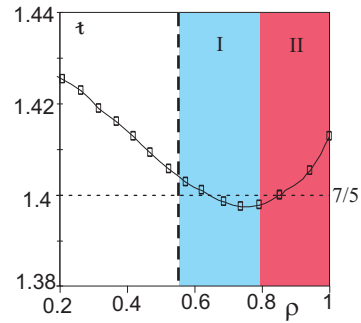}
\caption{\label{fig4} Radial profile of rotational transform in
magnetic configuration discussed in this paper.}
\end{figure}

The list of spatially coinciding phenomena in this plasma is
complemented by the observation that the TB region appears in the
vicinity of the zero magnetic shear in this magnetic
configuration. The computed radial profile of the rotational
transform $\iotabar = 1/q$ (where $q$ is the safety factor) is
shown in Fig.~\ref{fig4}. In addition to zero shear at $\rho
\approx$ 0.75, $\iotabar$ = 1.4 = $n/m$ = 7/5 rational surfaces
are present in both zones I (at $\rho \approx$ 0.65) and II (at
$\rho \approx$ 0.85). The accuracy of the $\iotabar$ computation
has been verified using experimental electron beam mapping
\cite{Shats_RSI_95}. The existence of the 7/5 rational surfaces is
also confirmed by the observation of the $m=5$ chain of magnetic
islands in the region of $\rho\approx (0.83-0.87)$. The plasma
current in the H-1 heliac is negligibly small ($\sim 10$ A) and
does not affect the vacuum magnetic structure.

A possible role of low-order rational surfaces in the formation of
H mode has been recognized since the first observation of H mode
in stellarators \cite{Wagner_PPCF_94,Ascasibar_PPCF_2002}. Spatial
correlation of the rational surfaces with the stationary zonal
flows, seen in Fig.~\ref{fig3}(b) and Fig.~\ref{fig4}, may be
indicative of the generation of stationary zonal flows due to the
influence of the rational surfaces, as suggested in
\cite{Hidalgo_PPCF_2001}.

\begin{figure}
\includegraphics{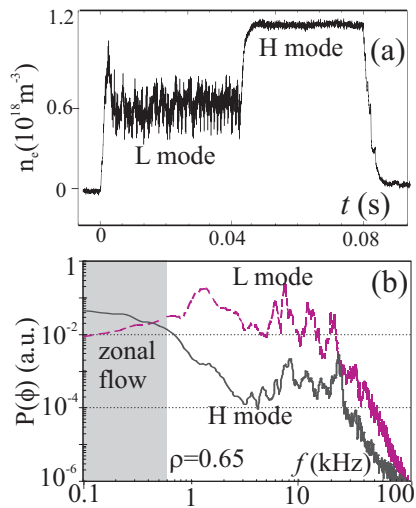}
\caption{\label{fig5}(a) Time evolution of the mean plasma density
during the spontaneous L-H transition, (b) power spectra of the
plasma floating potential across the L (dashed line) to H (solid
line) transition at $\rho=0.65$.}
\end{figure}

Formation of the TB and strong stationary zonal flow is also
observed during \textit{spontaneous} \mbox{L-H} transitions in
H-1, described in \cite{Punzmann2004}. Fig.~\ref{fig5}(a) shows
temporal evolution of the mean plasma density during spontaneous
L-H transition. In this discharge the mean electron density jumps
from about $0.6 \times 10^{18}$ m$^{-3}$ to almost $1.2 \times
10^{18}$ m$^{-3}$ in about one millisecond. Similarly to
stationary H mode discharges described above, strong zonal flow in
the TB region is observed in the H mode stage of discharges with
spontaneous transitions. The spatial correlation of the TB regions
and stationary zonal flow in the spontaneous transitions is also
observed.

In Fig.~\ref{fig5}(b), the change of the fluctuation power spectra
across the L-H transition at the radial position of $\rho=0.65$ is
illustrated. It can be seen that across the transition,
fluctuations in the broad spectral range from 0.6 to 100 kHz are
reduced, while the spectral power of low-frequency zonal flow, $f=
(0.1-0.6)$ kHz, is increased.

Many of the ingredients of the TB physics presented in this paper
have been discussed with regard to H modes in tokamaks and
stellarators. For example, the role of the low-order rational
surfaces \cite{Garcia_PoP_2001, Hahm_PPCF_2002}, the role of zonal
flows in L-H transition \cite{Fujisawa2006}, modification of the
$E_r$ profiles by non-neoclassical (turbulence-driven) effects
\cite{Diamond1991,Terry2000, Wagner2006}.

Here we present for the first time experimental evidence that
zonal flow, which develop near the $n/m$ = 7/5 rational surfaces,
is spatially correlated with distinct regions of the radial
electric field inside the TB region in the H mode plasma. The
resulting strong peaks in $E_r$ shear coincide with the kinks in
the density profile in H mode, suggesting that the strong peaks in
the $E_r^{\prime}$ define the position and the width of the TB.
Similar physical picture has recently emerged as a result of
analysis of gyrokinetic simulations of DIII-D tokamak discharges
\cite{Waltz2006}. Corrugations in the radial profiles of electron
density, temperature and radial electric field in DIII-D are
observed near low-order rational surfaces. The development of
strong zonal flows and strong $E_r$ shear layer in these plasma
regions suggests the development of zonal flow as a trigger for
the TB formation.

Since zonal flows are usually thought of as turbulence-driven
flows, possible mechanisms of the zonal flow enhancement and
sustainment in H mode, when the level of turbulence is
substantially reduced, need to be explained. It has been suggested
in \cite{Shats_PRE_05}, that the redistribution of spectral energy
from a broad range of intermediate scales into a stationary zonal
flow is the mechanism of the zonal flow enhancement during L-H
transitions. The result shown in Fig.~\ref{fig5}(b), to large
extent confirms this hypothesis in a radial region close to the
transport barrier. Also in \cite{Shats_PRE_05} we presented
experimental evidence that the nonlocal spectral transfer of
energy from the unstable drift wave is responsible for the
sustainment of zonal flow in H mode, as has been proposed in
\cite{Balk_JETP_90}. Further theoretical work in this direction is
needed.

\begin{acknowledgments}
The authors would like to thank Santhosh Kumar for providing
computed data on rotational transform.
\end{acknowledgments}

\end{document}